# The Corset Effect, Field-Cycling NMR Relaxometry, Transverse NMR Relaxation, Field-Gradient NMR Diffusometry, and Incoherent Neutron Scattering

Rainer Kimmich[1] and Nail Fatkullin[2]

[1]Universität Ulm, 89069 Ulm, Germany

[2]Department of Physics, Kazan State University, Kazan 420008, Russia, Tatarstan

The corset effect is an experimentally established dynamic confinement phenomenon first observed by field-cycling NMR relaxometry and transverse NMR relaxation in polymer melts in pores with dimensions ranging from a few nanometers up to 0.06 micrometers or even more. The techniques employed are specifically sensitive to rotational fluctuations of polymer segments. It will be shown that neutron scattering and other methods probing translational fluctuations are not suitable for the detection of this phenomenon.

The so-called corset effect was identified as a strong change of spin-lattice relaxation dispersion and a strong reduction of relaxation times in polymer melts confined in pores with dimensions ranging from nanometers even up to micrometers relative to bulk polymer melt samples.[1-7] This phenomenon indicates a strong slowing down of rotational polymer segment dynamics upon confinement. The measuring techniques were field-cycling NMR relaxometry and transverse relaxation.[8] The first study a decade ago where the corset effect manifested itself referred to a system of polyethyleneoxide strands embedded in a rigid cross-linked polyhydroxyethyl methacrylate (PHEMA) matrix.[1] The effect was reproduced later in different systems (polyethyleneoxide in variants of PHEMA, perfluoropolyether in Vycor and in layered structures), with different instruments (home built and STELAR Spinmaster FFC2000), and with different nuclei ($^1$H, $^2$H, and $^{19}$F). More recently an equivalent behavior was reported for immiscible polymer blends studied by mechanical relaxation.[9]

In a recent publication[10] it was concluded from neutron scattering data that the corset effect can be "ruled out". However, well established experimental phenomena are not negotiable. The discussion of the corset effect can only be a matter of whether the interpretation is adequate, and whether other techniques are appropriate for the detection of the phenomena. The present communication will therefore focus on three points for clarification: (A) the experimental evidence of the corset effect; (B) the sensitivity of different techniques to this particular phenomenon; (C) interpretations of the experimental findings.

*A) Experimental facts and evidence for the corset effect.* Fig. 1 shows typical field-cycling NMR relaxometry data sets demonstrating the corset effect.[1] Melts of linear perdeuterated polyethylene oxide (PEO) reveal much steeper dispersions of the deuteron spin-lattice relaxation time $T_1(\nu)$ when confined in a solid matrix (Fig. 1a) than in bulk (Fig.1b). The values of the spin-lattice and the transverse relaxation times as quantities reflecting rotational fluctuations of segments are strongly reduced. This is the corset effect: *Rotational segment dynamics is more constrained in confined polymers than in bulk.* Similar experiments with proton and fluorine NMR as well as with mechanical relaxation lead to equivalent results and have been described in Refs 3-9.

The matrix of the samples used for the data in Fig. 1a consisted of cross-linked polyhydroxyethyl methacrylate (PHEMA). The preparation of the nanoscopic PEO strands in this sample is based on spinodal decomposition and is described in detail in Refs 12 and 18. The strand diameter was estimated with different methods including electron microscopy.[11-13] The value for the samples studied in Ref. 1 is about 10 nm. Modifying the preparation protocol, samples with diameters in a range from 8 up to 60 nm can be produced.[12]

Remarkable features of the experiments represented by Fig. 1 are: (i) Deuteron resonance selectively probes the perdeuterated PEO strands; (ii) deuteron spin-lattice relaxation is exclusively sensitive to rotational fluctuations of the quadrupole spin interaction tensor, that is of the methylene segments of the polymer; (iii) the spin-lattice relaxation dispersion of the confined polymer approaches the power law $T_1 \propto M_w^0 \nu^{0.75}$, a feature of the tube/reptation model, at elevated frequencies in contrast to the bulk behavior; (iv) immaterial spin diffusion is not effective for deuteron spin-lattice relaxation and can therefore not act as an averaging mechanism.

Parallel to the field-cycling NMR relaxometry studies, translational diffusion measurements with the aid of the field-gradient NMR diffusometry technique were performed. The results were shown to be compatible with the tube/reptation model.[11,12] The dimension of the topological constraints concluded on this basis is however much larger than usually anticipated for this model and turned out to coincide with the real pore diameter. This means that no substantial confinement effect beyond restriction to the pores can be stated for translational diffusion. The strong effect found for spin-lattice relaxation in the form of the corset effect is obviously not effective for field-gradient NMR diffusometry.

Fig. 2 shows typical (normalized) data of stimulated-echo amplitudes attenuated by translational displacements. Actually, with respect to particle translations, the echo amplitudes are governed by a wave-number dependent function identical to the intermediate dynamic structure factor of incoherent neutron scattering (see below).[19] Apart from the intrinsically different time and wavenumber scales, conclusions based on these data should be essentially equivalent to those from neutron scattering data of confined samples. No really new information can therefore be expected on the basis of incoherent neutron scattering. An important condition is, however, that the same evaluation formalism is employed for both techniques. Envisaging the tube/reptation concept, the correct formalism is the specific treatment outlined in Refs 12,17.

*B) Discrimination of methods with respect to their sensitivity of rotational and translational fluctuations.* Molecular fluctuations are characterized by time autocorrelation functions. The correlation function probed by deuteron field-cycling NMR relaxometry in conjugate Fourier form is[8]

$$G_r(t) = \langle Y_{2,m}(t) Y_{2,-m}(0) \rangle \qquad (1)$$

where $Y_{2,m}(t)$ is a second-order spherical harmonic for $m = 1, 2$. Proton and fluorine spin-lattice relaxation are dominated by the same function modified by a usually negligible translational contribution. Transverse NMR relaxation also depends on this sort of correlation function with $m = 0, 1, 2$. The spherical harmonics in Eq. (1) is a function of the fluctuating polar and azimuthal angles $\vartheta(t)$ and $\varphi(t)$, respectively, of a molecular axis relative to the laboratory frame with the $z$ axis along the external magnetic field. In the present case, the molecular axis refers to a methylene group of PEO. That is, the correlation function Eq. (1) reflects the features of *rotational fluctuations* of the spin bearing entity. In the case of polymers, rotational fluctuations can essentially be taken to be synonym to conformational fluctuations.

The correlation function Eq. (1) is in contrast to the function commonly probed by incoherent neutron scattering[20] as well as by field-gradient NMR diffusometry[19]:

$$G_t = \langle e^{-i\mathbf{k}\cdot\mathbf{R}(0)} e^{i\mathbf{k}\cdot\mathbf{R}(t)} \rangle \qquad (2)$$

The position vector of a reference particle and the wave vector defined by the experimental set-up are represented by $\mathbf{R}$ and $\mathbf{k}$, respectively. This function clearly represents *translational fluctuations*. With respect to topological constraints of molecular dynamics, the information content is consequently different from that of Eq. (1). It may be noteworthy in the present context that coherent neutron scattering which probes displacements of different scattering centers *relative to each other* has also an NMR analogue in a sense: Evaluating the intermolecular dipolar interaction contribution to spin-lattice relaxation permits one to study relative diffusive displacements as demonstrated in Ref. 21. Thus both the "self" and "pair" correlation functions of incoherent and coherent neutron scattering, respectively, have NMR counterparts which should be considered in any study of translational fluctuations.

A very simple example illustrating the different manifestation of rotational and translational fluctuations applies to a nematic domain of a liquid crystal: Within the domain, translational fluctuations of molecules are unrestricted whereas rotational fluctuations are subject to substantial constraints. Another well-known case is molecular dynamics of fluids near surfaces: Relative to bulk fluids, there is a strong effect on rotational diffusion as probed by field-cycling NMR relaxometry.[22] On the other hand, the influence on translational diffusion is only minor. The latter finding was observed both by field-gradient NMR diffusometry[23] and neutron scattering.[24]

Of course, in an isotropic polymer melt translational and rotational fluctuations are largely but not entirely independent of each other (as it would be the case in a nematic liquid crystal domain). Some correlation must be expected. Then it is a matter of the accessible ranges of the experimental parameters whether such a correlation reveals itself in methods probing translational displacements. The time and length scales of neutron scattering and field-gradient NMR diffusometry are very different. While the incoherent neutron scattering technique used in Ref. 10 probes a time scale $t \leq 10^{-10}$ s and a length scale $8.3 \times 10^{-11}$ m $\leq k^{-1} \leq 5 \times 10^{-10}$ m = 0.5 nm, the field-gradient NMR diffusometry method employed in Ref. 12 is characterized by typical ranges $t \geq 10^{-2}$ s and $k^{-1} \geq 12.5$ nm. The respective limits of detectable root mean square particle displacements are $\langle r^2(t) \rangle^{1/2} < 0.7$ nm and $\langle r^2(t) \rangle^{1/2} > 10$ nm for incoherent neutron scattering and field-gradient NMR diffusometry. As will be seen below, the theories describing the corset effect predict a confinement dimension just in between so that both techniques are not expected to be sensitive to the aforementioned correlation. Note however, that the relatively long time and length scales intrinsic to field-gradient NMR diffusometry imply the scales on which the corset effect reveals itself. The NMR diffusometry decay curves (such as the ones shown in Fig. 2) can be described by the tube/reptation model with the electron microscopically determined pore diameter as the "tube" diameter *and bulk values* for all other parameters. This suggests that any influence of translation/rotation correlations must be minor.

*C) Interpretations and explanations.* The corset effect can be interpreted on different levels. The simplest conclusion one can draw from the different slopes of the $T_1$ dispersion in bulk and under confinement is a different degree of anisotropy of molecular dynamics. The steeper dispersion slope and the reduced transverse relaxation times under confinement mean that the decay of the correlation function Eq. (1) is retarded relative to the bulk sample. Segmental isotropization so-to-speak takes longer. The origin of this retardation can only be a stronger influence of topological constraints hindering rotational rearrangements of chain sections. The corset effect can therefore be traced back to more efficient topological constraints for conformational fluctuations.

On the next level, one can identify the corset effect as a finite-size phenomenon.[25] A well-known law of statistical physics relates the mean square fluctuation of the particle number in a given volume to the mean number of particles in that volume:

$$\left\langle \left(\delta N_V\right)^2 \right\rangle = k_B T \rho_s \kappa_T \left\langle N_V \right\rangle \tag{3}$$

$k_B$ is Boltzmann's constant, $T$ is the absolute temperature, $\rho_s$ is the mean (segment) particle number density, $\kappa_T$ is the isothermal compressibility of the system. That is, density fluctuations in large systems are larger than in small system. Reducing the system size by confinement consequently reduces the fluctuations. In terms of the fluctuating free volume, this means that conformational fluctuations are also restricted in their extent. The corresponding topological constraints are stronger. The analytical treatment in Refs 3, 25 suggest a dimension of those constraints in the order of

$$d \approx \sqrt{b^2 \rho_s k_B T \kappa_T} \tag{4}$$

where $b$ is the root mean square end-to-end distance of a statistical segment.

On the same basis, bulk behavior was estimated for the limit

$$d_{pore} \gg \left(\frac{b^3}{k_B T \kappa_T}\right)^{1/3} R_F \tag{5}$$

where $R_F$ is the Flory radius of a bulk polymer random coil. Inserting typical values for the parameters in Eqs (4) and (5) leads to values in the order of nanometers for the topological constraints causing the corset effect and in the order of micrometers for the crossover to bulk behavior.

The third level of interpretation refers to the Doi/Edwards model for reptation in a fictitious tube.[26] Identifying the extension of the relevant topological constraints with the fictitious Doi/Edwards tube unavoidably leads to a very tight diameter if the strong anisotropy of rotational fluctuations is to be explained. We note that the power law

$$T_1 \propto M_w^0 \nu^{0.75} \tag{6}$$

appearing in Fig. 1a in the upper half of the frequency window coincides with the prediction for the characteristic limit (II)$_{DE}$ of the tube/reptation model for a time scale between the so-called entanglement time and the longest Rouse relaxation time.[14] Interestingly the tube diameter derived from the experimental data on the basis of the tube/reptation model nicely fits to the value estimated on the basis of Eq. (4) when this expression is taken to represent the extension of the tube. As mentioned above, the conclusion suggested by this interpretation is valid specifically for rotational fluctuations. Translational fluctuations as revealed by field-gradient NMR diffusometry and neutron scattering cannot be very sensitive to this feature of molecular dynamics.

Finally, the role of surface adsorption as a potential origin of something like the corset effect should be considered. Actually this sort of phenomenon was suggested in Ref. 10 to be relevant for the samples under investigation there. Surface adsorption is a quite important issue that must be discussed for the specific polymer/matrix system examined. Let us now give four reasons why perceptible surface adsorption can be ruled out for the systems for which the corset effect was measured.

(i) From the experimental point of view, the existence of adsorbed, i.e. immobilized, phases would have revealed themselves as distinct relaxation or diffusion components (as it was actually observed in the polydimethylsiloxane/Vycor system studied in Ref. 27). This is not the case in the PHEMA samples and in the other systems for which a corset effect was identified, neither with field-cycling NMR relaxometry and transverse relaxation nor with field-gradient NMR diffusometry. As concerns spin-lattice relaxation, it should be recalled in this context that immaterial spin diffusion by flip-flop transitions as an averaging mechanism is not effective for deuterons. (ii) The PEO strands in methacrylate matrices were prepared on the basis of spinodal decomposition.[12,18] That is, the polymers under study were so-to-speak expelled from the matrix material, and softly repulsive polymer/wall interactions are expected. Likewise, the fluorine containing polymer studied in Ref. 4 is a compound with little affinity to the surfaces of the matrix material. (iii) In the case of the methacrylate samples, the PEO strand diameters and, hence, the surface to volume ratio were varied in a wide range with almost no effect on the results.[3] (iv) The corset effect vanishes when frequency/time scales are approached where short-range chain modes, so-called "local motions", dominate. The corset effect must therefore be characterized to be of a predominantly geometric nature not perceptibly affected by wall adsorption.

*Conclusions.* Neutron scattering and other methods probing translational fluctuations such as field-gradient NMR diffusometry are not sensitive to confinement effects due to topological constraints for rotational fluctuations. In the particular systems studied in Refs 1-7, there is no perceptible manifestation of surface adsorption. The corset effect can be explained on a geometrical basis as a finite- size confinement phenomenon.

**Acknowledgments**

**References**

1. R. Kimmich, R.-O. Seitter, U. Beginn, M. Möller, N. Fatkullin, Chem. Phys. Letters **307**, 147 (1999).
2. R. Kimmich, N. Fatkullin, R.-O. Seitter, E. Fischer, U. Beginn, M. Möller, Macromol. Symp. **146**, 109 (1999).
3. C. Mattea, N. Fatkullin, E. Fischer, U. Beginn, E. Anoardo, M. Kroutieva, R. Kimmich, Appl. Magn. Reson. **27**, 371 (2004).
4. R. Kausik, C. Mattea, R. Kimmich, N. Fatkullin, Eur. Phys. J. Special Topics **141**, 235 (2007).
5. N. Fatkullin, R. Kausik, R. Kimmich, J. Chem. Phys. **126**, 094904 (2007).
6. R. Kausik, N. Fatkullin, N. Hüsing, R. Kimmich, Magn. Reson. Imaging **25**, 489 (2007).
7. R. Kausik, C. Mattea, N. Fatkullin, R. Kimmich, J. Chem. Phys. **124**, 114903 (2006).
8. R. Kimmich, *NMR Tomography, Diffusometry, Relaxometry* (Springer, Berlin, 1997).
9. C.-Y. Liu, B. Zhang, J. He, R. Keunings, C. Bailly, Macromolecules **42**, 7982 (2009).
10. M. Kroutyeva, J. Martin, A. Arbe, J. Colmenero, C. Mijangos, G. J. Schneider, T. Unruh, Y. Su, D. Richter, J. Chem. Phys. **131**, 174901 (2009).
11. E. Fischer, R. Kimmich, U. Beginn, M. Möller, N. Fatkullin, Phys. Rev. E **59**, 4079 (1999).
12. E. Fischer, U. Beginn, N. Fatkullin, R. Kimmich, Macromolecules **37**, 3277 (2004).
13. E. Fischer, U. Beginn, N. Fatkullin, R. Kimmich, Magn. Reson. Imaging **23**, 379 (2005).
14. R. Kimmich, N. Fatkullin, Advan. Polym. Science **170**, 1 (2004).

[15] G. D. Smith, D. Y. Yoon, R. L. Jaffe, R. H. Colby, R. Krishnamoorti, L. J. Fetters, Macromolecules **29**, 3462 (1996).

[16] W. W. Graessley, S. F. Edwards, Polymer **22**, 1329 (1981).

[17] N. Fatkullin, R. Kimmich, Phys. Rev. E **52**, 3273 (1995).

[18] U. Beginn, E. Fischer, T. Pieper, F. Mellinger, R. Kimmich, M. Möller, J. Polym. Sci. A: Polym. Chem. **38**, 2041 (2000).

[19] G. Fleischer, F. Fujara, in: *NMR - Basic Principles and Progress, Solid State NMR*, ed. P. Diehl, E. Fluck, H. Günther, R. Kosfeld, J. Seelig, Vol. 30, Springer, Berlin, p. 159-207 (1994).

[20] D. Richter, M. Monkenbusch, A. Arbe, J. Comenero, Advan. Polym. Sci. **174**, 1 (2005).

[21] M. Kehr, N. Fatkullin, R. Kimmich, J. Chem. Phys. **126**, 094903 (2007).

[22] S. Stapf, R. Kimmich, R.-O. Seitter, Phys. Rev. Letters **75**, 2855 (1995).

[23] R. Kimmich, S. Stapf, A.I. Maklakov, V.D. Skirda, E.V. Khozina, Magn. Reson. Imaging **14**, 793 (1996).

[24] M.-C. Bellissent-Funel, K.F. Bradley, S.H. Chen, J. Lal, J. Teixeira, Physica A: Statistical Mechanics and its Applications **201**, 277 (1993).

[25] N. Fatkullin, R. Kimmich, E. Fischer, C. Mattea, U. Beginn, M. Kroutieva, New J. Physics **6**, 46 (2004).

[26] M. Doi, S. F. Edwards, *The Theory of Polymer Dynamics*, Oxford Univ. Pr., 1986.

[27] S. Stapf, R. Kimmich, Macromolecules **29**, 1638 (1996).

**Figure legends:**

**Fig. 1:** The corset effect demonstrated with the frequency dependence of the deuteron spin-lattice relaxation time of perdeuterated PEO confined in 10 nm pores of solid PHEMA at 80°C **(a)** and in bulk melts **(b)** [1,2]. The dispersion of the confined polymers verifies the law $T_1 \propto M_w^0 \nu^{0.75}$ at high frequencies as predicted for limit (II)$_{DE}$ of the tube/reptation model.[14] The low-frequency plateau observed with the confined polymers indicates that the correlation function implies components decaying more slowly than the magnetization relaxation curves, so that the Bloch/Wangsness/Redfield relaxation theory[8] is no longer valid in this regime. This is confirmed by the fact that the plateau value corresponds to the transverse relaxation time, $T_2$, for deuterons extrapolated from the high-field value measured at 9.4 T.

**Fig. 2:** Echo attenuation curves for polyethyleneoxide, $M_w$ = 11,200, confined to PHEMA pores at 80 °C as a function of the squared wave-number, $k^2$, for different diffusion times.[11,12] Anticipating the tube/reptation model and using the formalism developed for this particular model[12-17] the data can be described by the solid lines commonly fitted to the data. The tube diameter as the only fitting parameter was found to be $a = (8 \pm 1)$ nm for this sample. This value nicely coincides with the pore diameter estimated from electron micrographs of the same sample. Other parameter values such as the number of statistical segments per chain $N = M_w / 853$, the segmental diffusivity $D_0 = 9.66 \times 10^{-10}$ m$^2$/s and the root mean square end-to-end distance of a statistical segment $b = 8.38 \times 10^{-10}$ m were taken from the literature for – notabene – bulk melts.[15,16] The echo attenuation curves are based on the same intermediate dynamic structure factor as relevant for neutron scattering experiments.

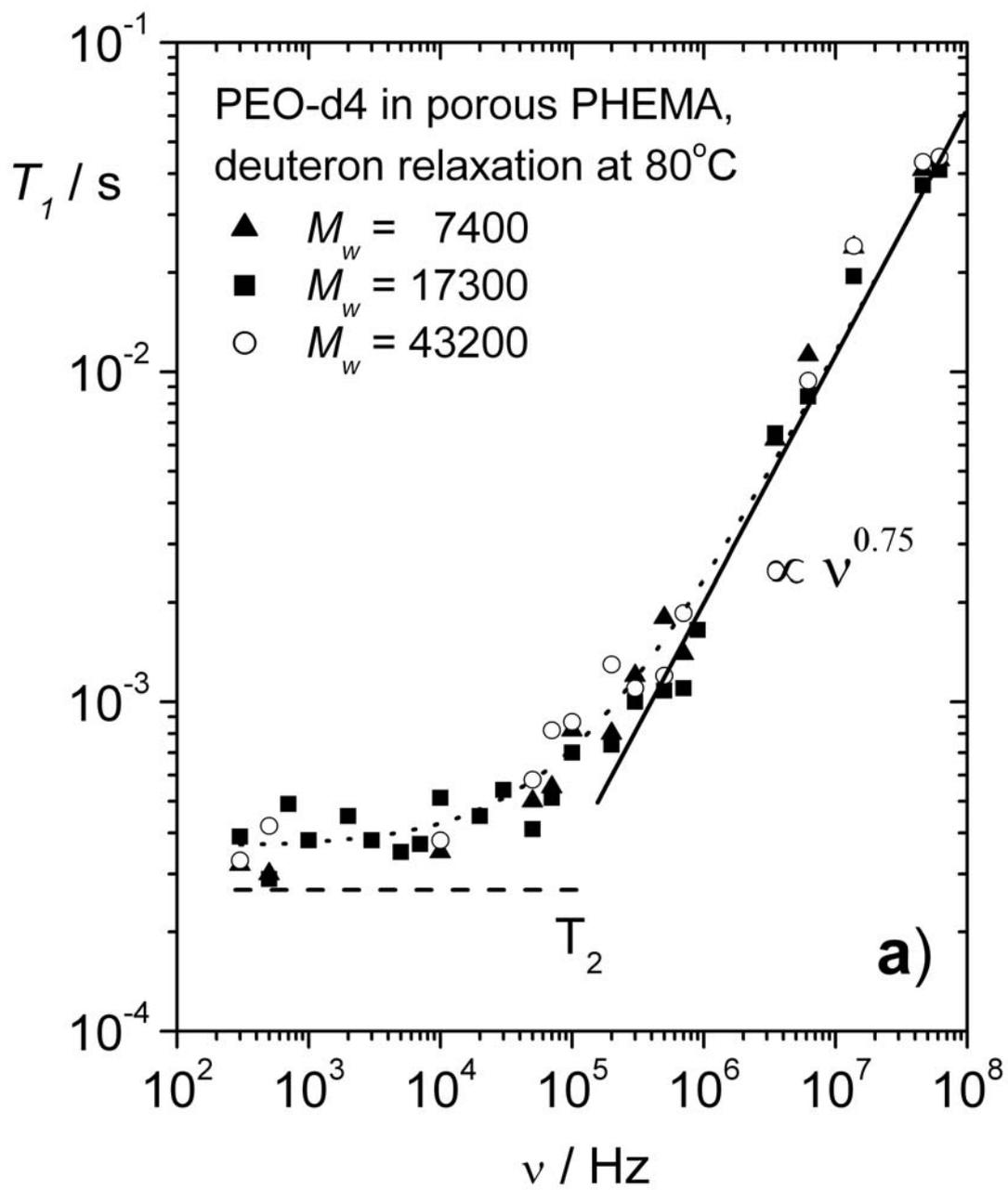

Fig. 1a

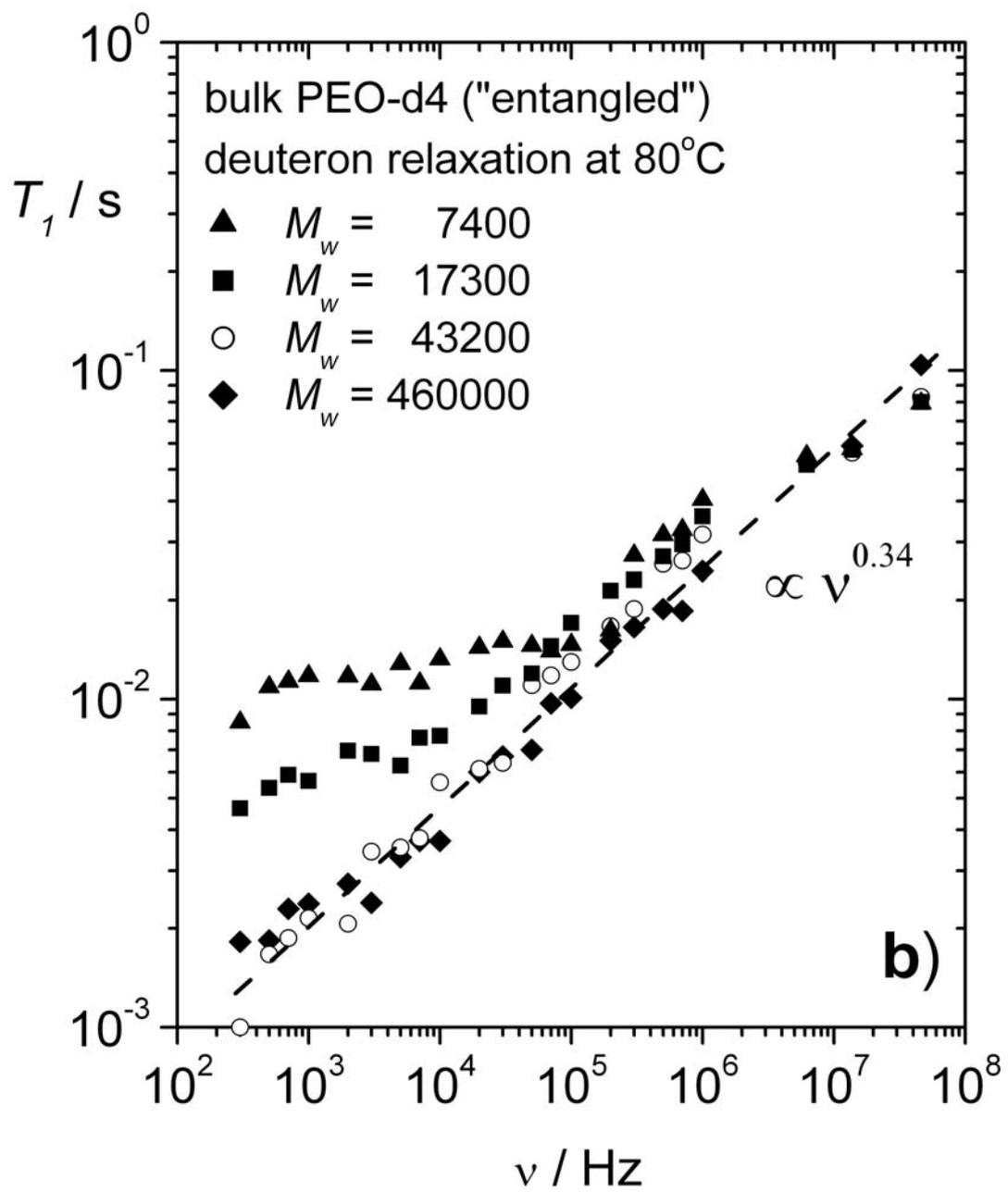

Fig. 1b

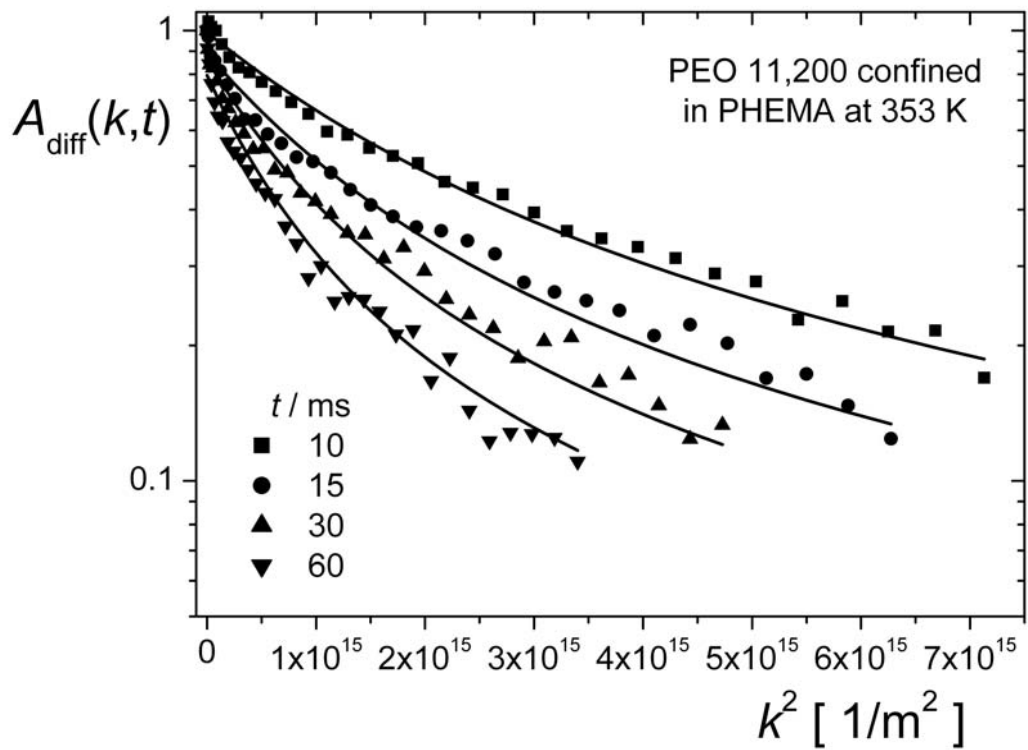

Fig. 2